 \documentstyle[aps,prc,floats,epsfig]{revtex}
 \draft
 \newcommand{\mm}{MM}
 \newcommand{\etal}{\emph{et al.}}
\begin{document}

\renewcommand{\thefootnote}{\arabic{footnote}}

\twocolumn[\columnwidth\textwidth\csname@twocolumnfalse\endcsname

\title{$\alpha$-decay chains 
       of $\bbox{{}^{289}_{175}114}$ and $\bbox{{}^{293}_{175}118}$ 
       in the relativistic mean-field model} 

\author{Michael Bender}

\address{Department of Physics and Astronomy, 
         University of Tennessee,
         Knoxville, Tennessee 37996
}

\address{Physics Division, 
         Oak Ridge National Laboratory,
         P.O.~Box 2008, Oak Ridge, Tennessee 37831
}

\address{Department of Physics and Astronomy,
         University of North Carolina, 
         Chapel Hill, North Carolina 27599
}

\date{September 9 1999}

\maketitle

\addvspace{5mm}

%
%
\begin{abstract}
A comparison of calculated and experimental $Q_\alpha$ values of 
superheavy even-even nuclei and a few selected odd-$N$ nuclei 
is presented in the framework of the relativistic mean-field model 
with the parameterization NL-Z2. Blocking effects are found to be 
important for a proper description of $Q_\alpha$ of odd mass nuclei.
The model gives a good overall description of the 
available experimental data. The mass and charge assignment of 
the recently measured decay chains from Dubna and Berkeley is in
agreement with the predictions of the model.
The analysis of the new data does not allow a 
final conclusion about the location of the expected island of 
spherical doubly-magic superheavy nuclei.
\end{abstract}


\addvspace{5mm}]

\narrowtext
%
%
In the last few years, the synthesis of superheavy nuclei with 
\mbox{$Z=110$}--$112$ at GSI (Darmstadt) and JINR Dubna -- see 
\cite{SHReview} and references therein -- has renewed the interest 
in the properties of superheavy nuclei. These are by definition 
the nuclei with \mbox{$Z > 100$} which have a negligible fission
barrier and are stabilized by shell effects only. The ultimate goal 
is to reach an expected 'island of stability' located around the next 
spherical doubly-magic nucleus which was predicted to be 
$^{298}_{184}114$ thirty years ago \cite{Nilsson,Mosel}.

Recently the discovery of new rather neutron-rich isotopes of the 
elements \mbox{$Z=112$} \cite{112171} and \mbox{$Z=114$} \cite{114173}
was reported from JINR Dubna, while an experiment at Berkeley led to 
the synthesis of even heavier nuclei with \mbox{$Z = 118$} \cite{118175}. 
While earlier superheavy nuclei could be unambiguously identified 
by their $\alpha$-decay chains leading to already known nuclei, the decay
chains of the new-found superheavy nuclei cannot be linked to any 
nuclei known so far. Their identification relies on comparison 
with theoretical models.

The experimental progress is accompanied by a large-scale investigation
of superheavy nuclei with the latest nuclear mean-field models. 
While refined macroscopic-microscopic (MM) models 
like the Yukawa-plus-exponential model with Woods-Saxon
single-particle potentials (YPE+WS) \cite{Smo97a} or the finite-range
droplet model with Folded-Yukawa single-particle potentials 
(FRDM+FY) \cite{Mol94a,FRDMval}
confirm the older prediction of $^{298}_{184}114$ for the next 
spherical doubly-magic nucleus, most self-consistent models shift 
that property to the more proton-rich $^{292}_{172}120$ 
or even $^{310}_{184}126$ \cite{Lal96,Cwi96,SHsphere,SHdef,SHshells}. 
The conflicting predictions have several reasons. \mm\ models have 
very restricted degrees of freedom of the radial density distribution
and the shape of the single-particle potentials, which becomes visible 
in superheavy nuclei where this hinders the occurrence of proton shells 
at \mbox{$Z=120$} or \mbox{$Z=126$} \cite{Cwi96,SHshells,Swi99}. 
Another source for different extrapolation to 
superheavy nuclei among the models is the spin-orbit interaction. 
While the self-consistent relativistic mean-field model (RMF) naturally
incorporates the nuclear spin-orbit interaction (which is a 
purely relativistic effect), it has to be put 
in by hand into all non-relativistic models, self-consistent 
ones using Skyrme (SHF) interactions and \mm\ models. Surprisingly
all non-relativistic models -- which have one or several additional 
parameters to explicitly adjust the experimental spin-orbit splittings 
-- perform not as good in this respect as the RMF which has no 
parameters adjusted to single-particle data at all. The non-relativistic 
models overestimate spin-orbit splittings with increasing mass 
number which might have some impact on their actual predictions
for shell closures in the superheavy region \cite{SHshells}.
Superheavy nuclei with their large density of single-particle 
states are a sensitive probe for models of nuclear shell structure
and can be used to discriminate among models which describe stable
nuclei with comparable quality.

Macroscopic-microscopic models -- which have been the favorite tool 
to describe superheavy nuclei for a long time -- provide a very good 
description of masses throughout the chart of nuclei, but they 
rely on preconceived knowledge about the densities and single-particle
potentials which fades away when going towards
the limits of stability. Although self-consistent models
have not yet reached the overall performance of \mm\ models, they
incorporate polarization effects on the density distribution
which might be crucial for a proper description of superheavy
nuclei where the strong Coulomb field pushes the protons to 
the nuclear surface \cite{Cwi96,SHsphere}.

Recently {\'C}wiok \etal\ have reported a detailed
comparison of the new experimental data with SHF calculations employing
the interaction SLy4 \cite{Cwi99}. Although the calculated
$Q_\alpha$ values show some deviations from the measured values, 
the analysis confirms the assignment of the mass and charge number
of the new nuclides. SLy4 shifts the island of stability 
towards very high charge numbers around ${}^{310}_{184}126$ 
\cite{Cwi96,SHsphere,SHdef,SHshells}. It is the aim of this
contribution to compare the experimental data with predictions
of the RMF model \cite{Rei89} obtained with the 
recent parameterization NL--Z2 by P.-G.\ Reinhard \cite{SHshells}.
This particular force provides the best overall description of
binding energies of superheavy nuclei among the
current parameterizations of the RMF \cite{SHdef,JRNproc}
and with ${}^{292}_{172}120$ gives an alternative prediction 
for the next doubly magic nucleus which is quite close to 
the upper end of the new \mbox{$Z = 118$} chain.
%
%
\begin{figure}[t!]
\centerline{\epsfig{file=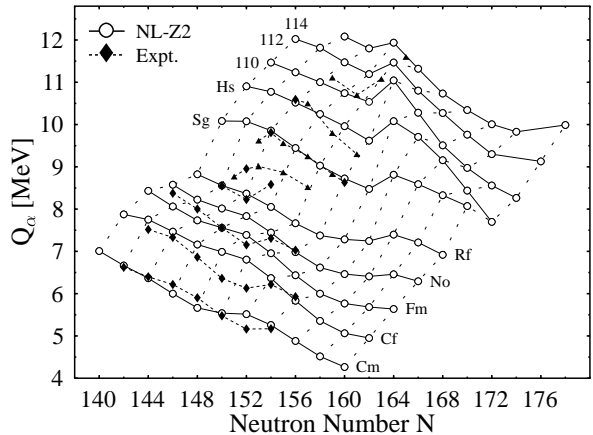}}
\caption{\label{Fig:Qalpha}
$Q_\alpha$ values of even-even nuclei in the superheavy region
calculated with NL-Z2 (open circles) compared with experimental
values for even-even nuclei (filled diamonds) where available
and odd-$N$ isotopes (open diamonds) of the heaviest even $Z$ 
elements. Solid lines connect nuclei with the same proton number 
$Z$, dotted lines nuclei in $\alpha$-decay chains. Including data
for nuclei with \mbox{$Z>116$} leads to overlapping curves, therefore
those are omitted in the plot.
}
\end{figure}
%
%

The calculations are performed on an axially symmetric 
grid assuming reflection symmetry.
Nuclei with odd mass number are calculated in a self-consistent
blocking approximation taking into account time-odd contributions
to the Dirac equation as described in \cite{oddNuclei}. 
Pairing correlations are treated within the BCS scheme using a 
delta pairing force, see \cite{SHshells} for details.
There remains a small uncertainty of the 
calculated $Q_\alpha$ values on the order 
of a few 100 keV from a possible variation of pairing recipes 
(i.e.\ particle-number projection and choice of pairing interaction), 
especially in odd nuclei. The pairing 
strength is fitted to pairing gaps calculated
in spherical blocking approximation. Taking polarization effects
in odd nuclei into account during the fit gives larger pairing strengths 
\cite{xu,uggap}, but this affects mainly the odd-even staggering, while
the $Q_\alpha$ of heavy nuclei are rather robust. The correction for 
spurious center-of-mass motion is performed a posteriori
as done in the original fit of NL--Z2, while corrections for spurious 
rotational and vibrational motions are neglected (as it is done in all 
other studies of superheavy nuclei) which might affect $Q_\alpha$ 
values in some cases to the order of 1 MeV and tend to diminish
shell effects \cite{Corr}.

A first test of the reliability of the RMF to describe the new
data is to check its performance for known $Q_\alpha$ values
of even-even nuclei, see Figure~\ref{Fig:Qalpha}. 
Owing to the lack of data for even-even nuclei the $Q_\alpha$ of 
odd-$N$ nuclei are drawn beyond \mbox{$Z=104$}.
They have to be handled very carefully, some of these 
$Q_\alpha$ values might correspond to transitions involving 
excited states, and due to the blocking effects the 
ground-state-to-ground state $Q_\alpha$ values might deviate on 
the order of 500 keV from the systematics of the $Q_\alpha$ values of
even-even nuclei. NL-Z2 gives a good overall description of the 
data except for some nuclei around Rf$_{104}$ where NL-Z2 overestimates 
a deformed proton sub-shell closure while the deformed \mbox{$N=152$} 
shell is underestimated. The latter is a problem from which virtually 
all self-consistent models suffer \cite{SHdef,Cwi99}. 
The particular Skyrme interaction SLy4 used in \cite{Cwi99} 
performs better in the region around \mbox{$Z=104$}, but 
it is to be noted that this interaction has 
small errors in the isotopic and isotonic mass systematics of 
these nuclei \cite{SHdef,JRNproc}, which cancel when calculating 
$Q_\alpha$ values.
%
%
\begin{figure}[b!]
\centerline{\epsfig{file=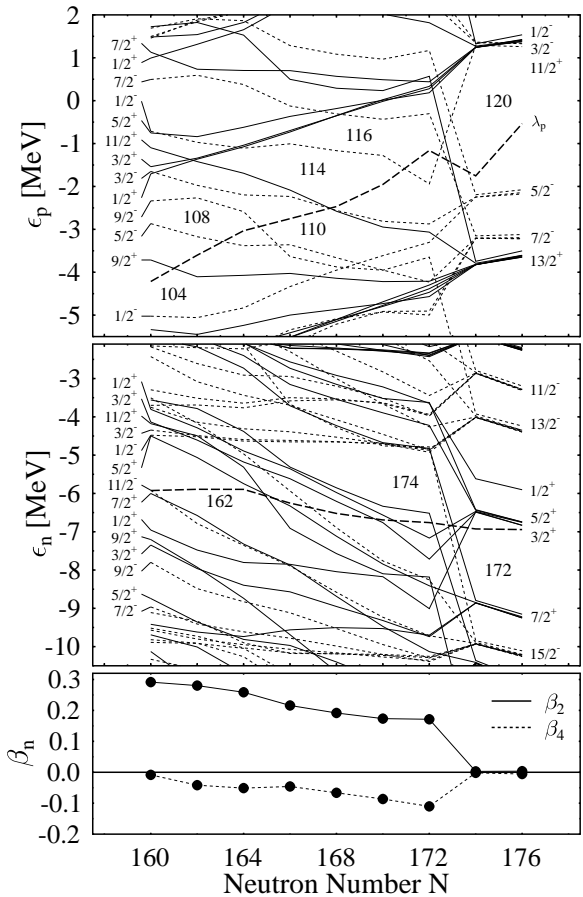}}
\caption{\label{Fig:174_118_spect}
Single-particle spectrum of the protons (upper panel), neutrons
(middle panel) and the relative quadrupole and hexadecapole moments
given by $\beta_\ell = 4 \pi \langle r^2 Y_{\ell 0} \rangle/(3 A R^\ell)$ 
with $R = 1.2 \, A^{1/3}$ fm (lower panel) 
for the $\alpha$-decay chain ${}^{296}_{176}120 \rightarrow  
{}^{292}_{174}118 \rightarrow \cdots \rightarrow 
{}^{264}_{160}{\rm Rf}_{104}$ of even-even nuclei 
adjacent to the $\alpha$-decay chain of $^{293}_{175}{118}$.
For the deformed nuclei at small neutron number the states are 
labeled with angular-momentum projection and parity, while for the
spherical nuclei at large neutron number the total angular momentum 
and parity are given. The full (dotted) lines in the upper and middle 
panels denote positive (negative) parity states, the dashed lines 
the Fermi energies.
}
\end{figure}
%
%

To get an impression of the shell structure (as predicted 
by NL-Z2) of nuclei in the new-found region,
Fig.~\ref{Fig:174_118_spect} shows the 
single-particle spectra of the even-even nuclei with one
neutron less than those in the decay chain of ${}^{293}_{175}{118}$. 
The nuclei close to the \mbox{$Z=120$} shell closure 
come out spherical including ${}^{292}_{174}{118}$, while 
all lighter nuclei have deformed ground states. In addition to 
the spherical \mbox{$Z=120$} shell, the proton spectra show several
deformed shell closures indicated in the plot. 
In contrast to \mm\ models and some SHF interactions, however, the proton
spectra show no preferred (deformed) shell closure for the light 
nuclei in this chain, but a region of small level density between
\mbox{$Z=104$} and \mbox{$Z=110$}.  
The single-particle spectra of the neutrons are much denser with 
a spherical \mbox{$N=172$} shell and 
deformed shell closures at \mbox{$N=174$} and \mbox{$N=162$}.
The latter ones are also predicted by other
models \cite{Smo97a,Mol94a,Cwi99} and \mbox{$N=162$} is
already confirmed experimentally \cite{SHReview}.
The single-particle spectra of nuclei close to the decay chain 
of $^{289}_{175}{114}$ look similar except for the fact that 
all nuclei there are deformed.

\begin{figure}[t]
\centerline{\epsfig{file=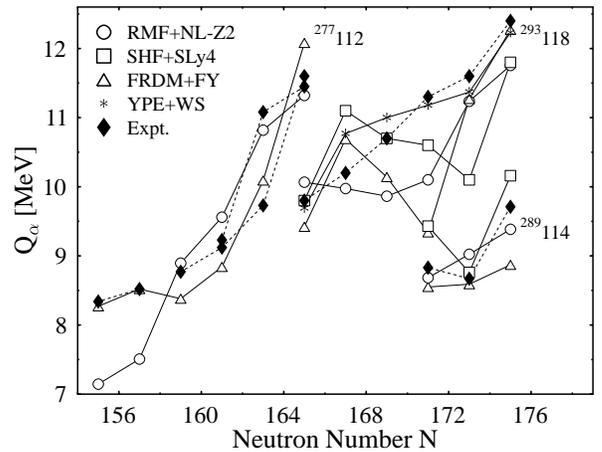}}
\caption{\label{Fig:ug:Q:alpha}
Comparison of experimental and calculated $Q_\alpha$ values
for the decay chains of $^{277}_{165}112$, $^{289}_{175}114$,
and $^{293}_{175}118$, in the latter two cases following the
mass and charge assignment of the experimental groups.  
In the $^{277}_{165}112$ chain two 
distinct branches leading through different states of
the intermediate nuclei are known. The calculated values 
from RMF+NL--Z2 and SHF+SLy4 connect the lowest 
states with positive parity in all cases (in the new chains 
only ${}^{289}_{175}114$ and ${}^{277}_{167}110$ are predicted 
by NL-Z2 to have ground states with negative parity in agreement 
with \protect\cite{Cwi99}), while the FRDM+FY and YPE+WS data 
are ground state to ground state values.
}
\end{figure}

Three $\alpha$ decay chains of odd--mass nuclei are discussed 
in the following: the one starting with ${}^{277}_{165}112$ measured at
GSI in 1994 \cite{165112} serving as a testing ground for the
performance of NL-Z2 when describing odd-mass number 
nuclei, and the new data assigned to ${}^{289}_{175}114$ 
and ${}^{293}_{175}118$. The experimental data were obtained from two,
one and three events respectively.
Fig~\ref{Fig:ug:Q:alpha} compares calculated and measured $Q_\alpha$ 
values of these selected chains. Values calculated with the
FRDM+FY model \cite{FRDMval}, the YPE+WS model \cite{Smo99}
and the SHF+SLy4 model \cite{Cwi99} are given for comparison.
As can be seen from Fig.~\ref{Fig:174_118_spect}, the quasiparticle 
spectra of odd-$N$ nuclei in this region are very dense. Each 
single-quasiparticle state is the band head of a rotational band 
with transition energies from the first exited state to the 
band head below 50 keV which 
further increases the level density significantly. It cannot be 
expected that the synthesis of these nuclei and their decay chains 
lead always to ground states but very often to excited states.
Experimental evidence for that was found in the decay of ${}^{277}_{165}112$,  
each of the two nuclei identified so far \cite{165112} has
decayed through a different state in ${}^{273}_{163}110$. The
branch with large $Q_\alpha$ in ${}^{273}_{163}110$ presumably  
connects low-lying states in both nuclei while the branch with the 
low $Q_\alpha$ goes through a highly excited 
state in that nucleus. This demonstrates a fundamental problem 
when comparing measured and calculated $Q_\alpha$ of decay chains
with a few identified events only: the low statistics does not allow
to identify the whole spectrum of possible $\alpha$ transitions.
Detailed information on the $\alpha$-decay fine structure of
superheavy nuclei is available for a few nuclides up to \mbox{$Z=104$} 
only \cite{Hes97a}.
Some guidance when transitions to or between excited states might 
be favored is given by the fact that among transitions with similar
$Q_\alpha$ values those between states with the same quantum numbers 
are favored.
%
%
%
%

In view of the remaining uncertainties
NL-Z2 gives a very good description of the heavy nuclei in the decay chain
of ${}^{277}_{165}112$ above \mbox{$N > 157$} and reproduces the large jump 
in $Q_\alpha$ between ${}^{273}_{163}110$ and ${}^{269}_{161}108$ 
caused by the deformed \mbox{$N=162$} shell closure, which cannot be
seen in the predictions of the FRDM+FY model. For small neutron 
numbers the $Q_\alpha$ calculated with NL-Z2 are too small as in 
the case of the even-even nuclei discussed above. 
NL-Z2 gives also a rather good description of the
$\alpha$-decay chain of ${}^{289}_{175}114$. The missing kink
in the calculated values can be explained assuming that the 
decay of ${}^{289}_{175}114$ leads to one of the numerous low-lying 
excited states in ${}^{285}_{173}112$.

The decay chain of ${}^{293}_{175}118$ leads through a region with 
seemingly constant total shell effects, while most mean-field models
predict several spherical or deformed shell closures in this 
region. The FRDM+FY data show a pronounced shell effect for 
${}^{285}_{171}114$, while the SHF+SLy4 results
show one for ${}^{289}_{173}116$ which 
is related to deformed shells at \mbox{$Z=116$} and \mbox{$N=174$}
\cite{Cwi99}. NL-Z2 predicts a deformed \mbox{$N=174$} shell as 
well but a deformed \mbox{$Z=114$} proton shell 
(see Fig.~\ref{Fig:174_118_spect}) that leads to a broad plateau 
at smaller atomic numbers.
The predictions of the YPE+WS model of \cite{Smo97a,Smo99} 
follow qualitatively the experimental data for the heavy nuclei 
in this chain but show a shell effect for ${}^{277}_{167}110$. 
In contrast to the other models discussed here
the YPE+WS model predicts the nuclei at the upper end of this decay 
chain to be oblate deformed \cite{Smo97a}.

%
%
\begin{figure}[t]
\centerline{\epsfig{file=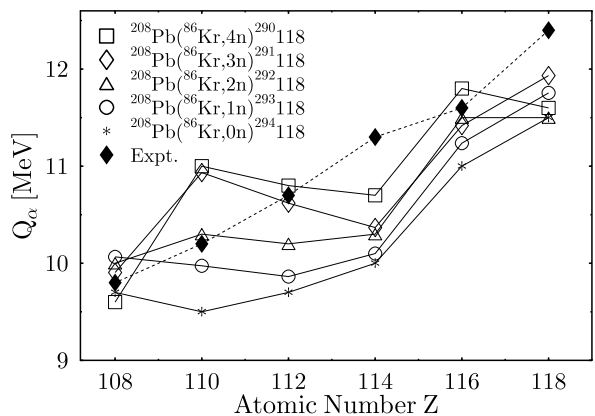}}
\caption{\label{Fig:ug:Q:alpha:2}
Comparison of experimental and calculated $Q_\alpha$ values
for the $\alpha$ decay chain of the $Z=118$ nuclei recently
measured in Berkeley assuming other neutron numbers of the initial 
nuclei. The $Q_\alpha$ values connect the lowest states with 
positive parity.
}
\end{figure}
%
%

The mass and charge assignment of the Dubna and Berkeley data
are based on theoretical models. While the proton number is
rather certain in both cases, the neutron number $N$ might also be smaller. 
Fig~\ref{Fig:ug:Q:alpha:2} shows the $Q_\alpha$ values measured in Berkeley
in comparison with calculated $Q_\alpha$ values for decay chains with 
varied neutron number of the initial nucleus. The heavy nuclei in
all decay chains have very similar $Q_\alpha$ values, while there are 
visible differences for the lower end of the decay chain. It can 
be clearly seen that $Q_\alpha$ of odd nuclei do not necessarily 
follow exactly the trend of the $Q_\alpha$ of the adjacent 
even-even nuclei. Since the $2n$ and higher reaction channels are
energetically forbidden and statistically suppressed, an
interpretation of the Berkeley data in terms of the $\alpha$-decay 
chain of ${}^{175}_{293}118$ is in agreement with predictions 
of the RMF with NL-Z2.
An interesting feature of the NL-Z2 results is
that $Q_\alpha$ values calculated between the lowest 
negative-parity quasiparticle states starting with ${}^{173}_{291}118$ 
or ${}^{175}_{293}118$ follow very closely the experimental data, but 
several of the intermediate nuclei can be expected to $\gamma$ decay to 
states in rotational bands build on lower-lying quasiparticle states 
with positive parity which prevents a long $\alpha$-decay chain between
negative-parity states (see also \cite{Cwi99}).

To summarize: The RMF with the parameterization NL-Z2 gives a reasonable
description of the $Q_\alpha$ values of known superheavy nuclei with
an overall quality comparable to other mean-field models that 
predict different spherical magic numbers.
On the basis of the available experimental data, a decision about
the location of the 'island of stability' cannot be made. The analysis
of the new data from Dubna and Berkeley in terms of the RMF 
model suggests that the relatively small $Q_\alpha$ values and 
corresponding long half-lives are caused by \emph{deformed} 
\mbox{$Z=114$} and \mbox{$N=174$} shell closures rather than the 
vicinity of the (potential) doubly magic ${}^{298}_{184}114$ predicted
by some other models, but 
a \emph{spherical} \mbox{$Z=114$} shell, restricted to higher  
neutron numbers, cannot be excluded. The spherical shell closures at 
\mbox{$Z=120$} and \mbox{$N=172$} predicted by the RMF are relatively
weak and do not significantly change the systematics of $Q_\alpha$
values of odd-$N$ nuclei at the upper end of the \mbox{$Z=118$} 
chain measured in Berkeley.
%
%

The author thanks W.\ Nazarewicz, V.\ Ninov, and G.\ M\"unzenberg
for inspiring discussions and communication of their results prior 
to publication, P.-G.\ Reinhard, K.\ Rutz and J.\ A.\ Maruhn for 
many fruitful discussions, and K. Rutz additionally for the RMF code 
used in this study. This work was supported by the U.S.\ Department 
of Energy under Contract Nos.\ 
DE-FG02-96ER40963 (University of Tennessee),
DE-FG02-97ER41019 (University of North Carolina), and
DE-AC05-96OR22464 with Lockheed Martin Energy Research Corp.\ 
(Oak Ridge National Laboratory).
%
%

%
%

\end{document}